\def\year{2021}\relax
\documentclass[letterpaper]{article} 
\usepackage{aaai21}  
\usepackage{times}  
\usepackage{helvet} 
\usepackage{courier}  
\usepackage[hyphens]{url}  
\usepackage{graphicx} 
\usepackage{booktabs}
\usepackage{natbib}  
\usepackage{caption} 
\usepackage{listings}
\usepackage{xcolor}
\definecolor{dkred}{rgb}{0.5,0,0}
\definecolor{dkgreen}{rgb}{0,0.6,0}
\definecolor{gray}{rgb}{0.5,0.5,0.5}
\definecolor{mauve}{rgb}{0.58,0,0.82}

\title{TODS: An Automated Time Series Outlier Detection System}
\author{
    Kwei-Herng Lai$^1$\thanks{Those authors contribute equally to this project}, Daochen Zha$^1$\footnotemark[1], Guanchu Wang$^1$, Junjie Xu$^1$, Yue Zhao$^2$, Devesh Kumar$^1$, Yile Chen$^1$, Purav Zumkhawaka$^1$, Mingyang Wan$^1$, Diego Martinez$^1$, Xia Hu$^1$\\
}
\affiliations{
    $^1$Department of Computer Science and Engineering, Texas A\&M University\\
    $^2$H. John Heinz III College, Carnegie Mellon University\\

    \{khlai,daochen.zha,hegsns,devesh,yilechen17,puravzum,w1996,dmartinez05,xiahu\}@tamu.edu,\\ xu.jj724@gmail.com, zhaoy@cmu.edu

}

\begin{document}

\maketitle

\begin{abstract}
We present TODS, an automated \textbf{T}ime Series \textbf{O}utlier \textbf{D}etection \textbf{S}ystem for research and industrial applications. TODS is a highly modular system that supports easy pipeline construction. The basic building block of TODS is primitive, which is an implementation of a function with hyperparameters. TODS currently supports 70 primitives, including data processing, time series processing, feature analysis, detection algorithms, and a reinforcement module. Users can freely construct a pipeline using these primitives and perform end-to-end outlier detection with the constructed pipeline. TODS provides a Graphical User Interface (GUI), where users can flexibly design a pipeline with drag-and-drop. Moreover, a data-driven searcher is provided to automatically discover the most suitable pipelines given a dataset. TODS is released under Apache 2.0 license at \url{https://github.com/datamllab/tods}. A video is available on YouTube\footnote{YouTube: \url{https://rb.gy/th5pcn}}.

\end{abstract}

\section{Introduction}
Outlier detection aims to identify unexpected or rare instances in data. As one of the most important tasks of data analysis, outlier detection has various real-world applications. For example, Yahoo~\cite{laptev2015generic} and Microsoft~\cite{ren2019time} have built their own time-series outlier detection services to monitor their business data and trigger alerts for outliers.

While machine learning models have shown promise in time series outlier detection, building an effective time series outlier detection pipeline heavily relies on human expertise. Due to the unique data characteristics of different application scenarios, one often needs expensive laboring trials to implement and identify the most suitable processing modules, detection algorithms, and hyperparameters for a specific task, which significantly impedes the real-world application of time series outlier detection.

\begin{figure}
    \centering
    \includegraphics[width=0.9\linewidth]{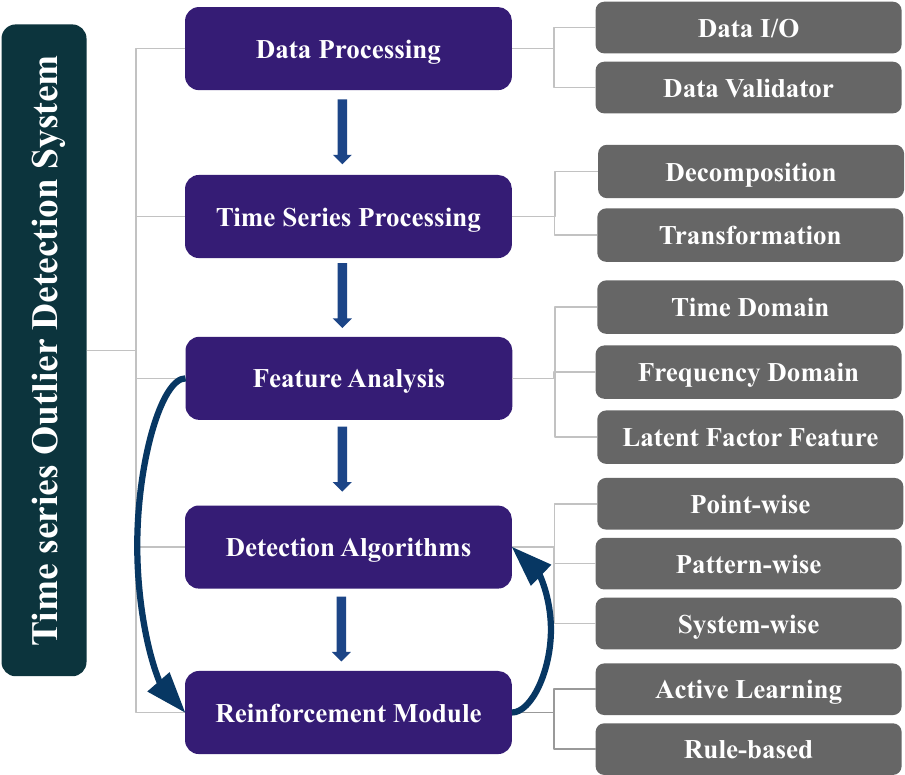}
    \caption{System overview. TODS provides end-to-end pipeline construction from data processing to detection algorithms. It also provides a reinforcement module to incorporate human knowledge into predictions.}
    \label{fig:overview}
\vspace{-0.55cm}
\end{figure}

To bridge this gap, we aim to save human efforts to build effective pipelines for time series outlier detection. Specifically, we implement commonly used time series analysis modules and detection algorithms under unified interfaces so that users can easily and flexibly try different pipelines. In addition, we provide data-driven searchers to help the users to automatically discover suitable pipelines. While our previous work has delivered an open-source toolbox for outlier detection~\cite{zhao2019pyod} and explored automated machine learning for outlier detection~\cite{li2020pyodds,li2020autood}, they provide limited automation for pipeline construction and are not designed for time series data. Whereas, we provide a highly modular system with a wide range of algorithm choices, specifically designed for various time series outlier detection scenarios. 

We present an automated \textbf{T}ime series \textbf{O}utlier \textbf{D}etection \textbf{S}ystem called \textbf{TODS}. It is designed under D3M infrastructure~\cite{milutinovic2020on}, i.e., data-driven model discovery via automated machine learning.
It currently supports 70 primitives for data processing, time series processing, feature analysis, outlier detection, and incorporating human knowledge. It can be applied to various application scenarios, including point-wise, pattern-wise and system-wise detection. The goal of TODS is providing an end-to-end solution for real-world time series outlier detection.


\begin{table}[]
  
  \centering
  \small
  \label{tab:dataset}
  \begin{tabular}{l|cc}
    \toprule
     Module & Number & Examples\\
    \midrule
    \midrule
    Data Processing & 7 & TimestampValidation \\
    Time Series Processing & 9 & SeasonalityDecomposition \\
    Feature Analysis & 30 & AutoCorrelation, NMF  \\
    Detection Algorithms & 23 & LSTMOD, IForest \\
    Reinforcement Module & 1 & RuleBasedFilter \\
    \midrule
    Total & 70 & - \\
    
  \bottomrule
\end{tabular}
\caption{The number of primitives implemented in each module of TODS. We only list some example algorithms in each module due to space limitation.}
\label{tbl:1}
\vspace{-10pt}
\end{table}

\begin{lstlisting}[title={Example 1: Code Snippet of function calls.},captionpos=b]
from tods import generate_dataset
from tods import load_pipeline
from tods import evaluate_pipeline

# Read data and generate dataset
df = pd.read_csv(table_path)
dataset = generate_dataset(df, target_index)

# Load pipeline
pipeline = load_pipeline(pipeline_path)

# Run the pipeline
evaluate_pipeline(dataset, pipeline, metric)
  
\end{lstlisting}
\vspace{-12pt}





\section{TODS System}

An overview of TODS is shown in Figure~\ref{fig:overview}. The basic building block in TODS is \emph{primitive},  which is an implementation of a function with some hyperparameters (e.g., Auto-Encoder algorithm). A \emph{pipeline} is defined as a Directed Acyclic Graph (DAG), where each step represents a primitive. A typical pipeline has four primitive steps, including data processing, time series processing, feature analysis, and detection algorithms. An optional reinforcement primitive can be added to the end of a pipeline to incorporate pre-defined rules. The primitives developed in each module are shown in Table~\ref{tbl:1}. Various pipelines can be constructed based on these primitives in different application scenarios. To enable easy pipeline construction, we provide a Graphical User Interface (GUI) to build and evaluate pipelines with drag-and-drop. In addition, a data-driven searcher is provided to automatically discover the suitable pipelines for a given task.

\subsection{Interface Design}
The interface is built upon Axolotl\footnote{\url{https://gitlab.com/axolotl1/axolotl}}, which is a high-level abstraction of D3M developed by our team. The pipeline is described with pipeline language. More details about pipeline language can be found in~\cite{milutinovic2020on}. To run a pipeline, we take a \texttt{CSV} file, a target index, and a pipeline description file as input, where the \texttt{CSV} file consists of timestamps, several features, and a target. A minimum example of evaluating a pipeline is shown in Example~1, where \texttt{pipeline\_path} is the path of pipeline description, \texttt{table\_path} is the path of \texttt{CSV} file, and \texttt{target\_index} describes which column is the target.

\begin{figure}
    \centering
    \includegraphics[width=1.0\linewidth]{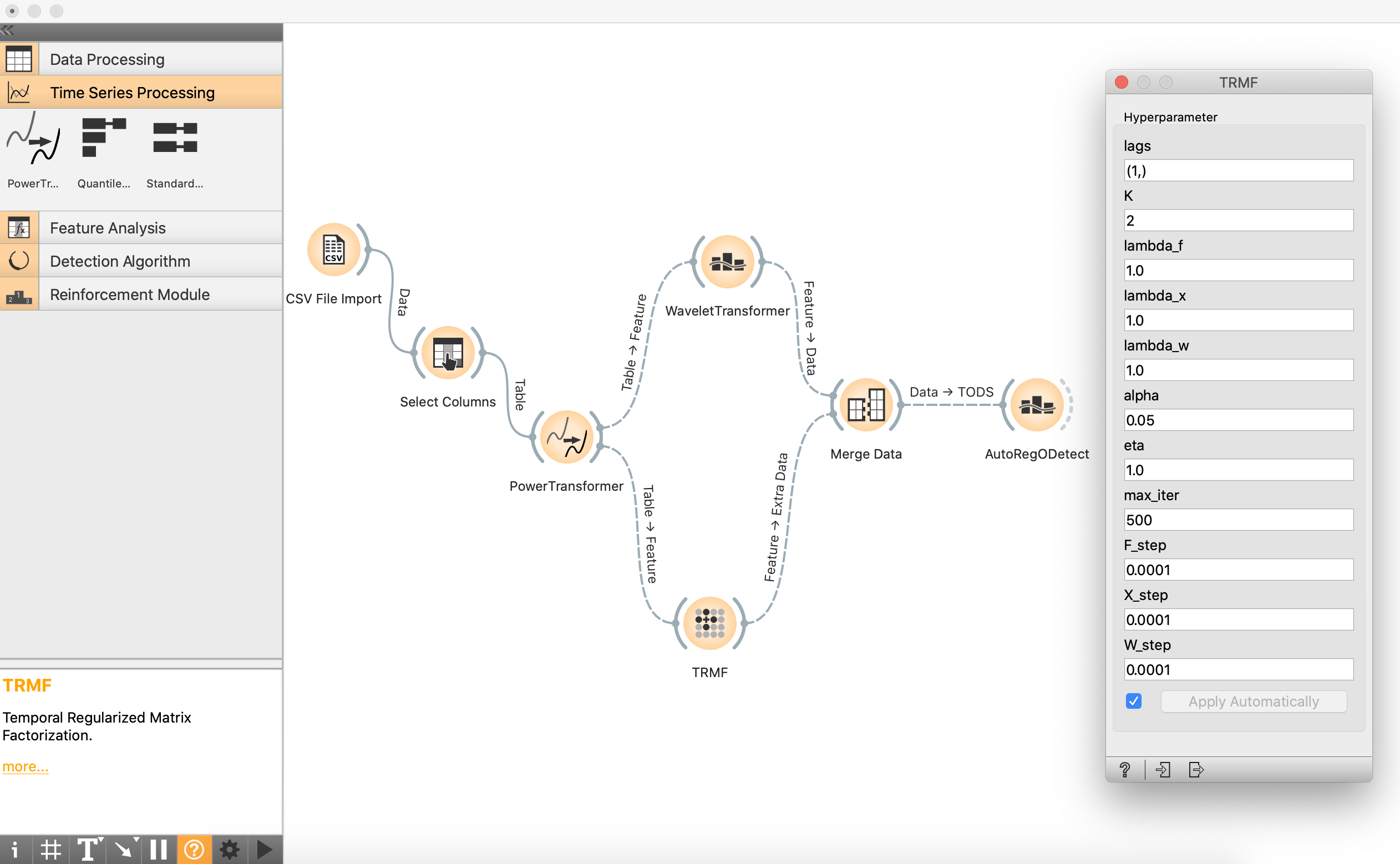}
    \caption{An illustration of the GUI. Users can easily build and evaluate a pipeline with drag-and-drop.}
    \label{fig:gui}
\end{figure}

\subsection{Live and Interective Part}
In the demo, we will show the Graphical User Interface~(GUI) of TODS. The GUI is implemented based on Orange~\cite{demvsar2004orange}. Figure~\ref{fig:gui} shows an example of how users can build a pipeline with GUI. The primitives in TODS are categorized into several tabs. Each primitive will have a corresponding icon and a text description. Users can move the primitives of interest to the canvas and connect them with lines. Users can also easily configure the hyperparameters of each primitive by double-clicking the icon, as shown in the pop-up window in the figure. TODS will automatically generate the pipeline language based on the structure in GUI. The generated pipeline will then be passed to the backend to perform time series outlier detection.

\subsection{Automated Pipeline Discovery}
Beyond manual pipeline construction, TODS provides data-driven searchers to automatically find good pipelines for a given task. In addition to the detection pipeline, TODS provides a data preparation pipeline for data splitting, such as k-fold splitting, and a scoring pipeline to evaluate the performance based on a metric, such as F1 score. The data-driven searcher takes a dataset as input to automatically discover a detection pipeline that can achieve the best performance on a given dataset. Then the discovered pipeline can be outputted for real-world deployment.

\section{Conclusions and Future Work}
TODS is a highly modular open-source system for automated time series outlier detection. We build TODS upon our previous research and open-source efforts in automated machine learning and outlier detection, hoping that it can enable end-users to quickly deploy a pipeline for a specific application scenario. In the future, we will add more primitives and design more efficient searchers for pipeline discovery. We will also improve the reinforcement module with learning-based active learning~\cite{zha2020metaaad}.

\bibliography{ref}

\end{document}